# Where's the Line? A Classroom Activity on Ethical and Constructive Use of Generative AI in Physics

Zosia Krusberg[1]
The University of Chicago

## Abstract

Generative AI tools like ChatGPT are rapidly reshaping how students and instructors engage with course material—and how they think about academic integrity. This paper presents a classroom activity designed to help physics students critically examine the ethical and educational implications of using AI in coursework. Through a structured sequence of scenario analysis, boundary-setting, and reflective discussion, with optional individual policy writing, students develop the metacognitive, ethical, and collaborative capacities needed to navigate emerging technologies thoughtfully and responsibly. Grounded in research on social constructivist learning, metacognition, and ethics education, the activity positions students as co-creators of an engaged and reflective learning environment.

[1]Email: zosia@uchicago.edu

## The Case for Ethical AI Engagement

The sudden emergence of generative AI in higher education—and the rapid pace at which the technology continues to evolve—has left many institutions scrambling to respond. In a short period of time, tools like ChatGPT moved from novelty to ubiquity, raising urgent questions about academic integrity, authorship, and the role of AI in learning. Institutional responses have ranged from blanket bans to open-ended permissiveness, often with limited guidance for students. Few approaches invite learners themselves to define what responsible AI use might look like. The activity described here offers a more participatory model—one that centers student reasoning and cultivates ethical engagement through dialogue and collaborative inquiry.

This approach also addresses a gap in the physics education literature. A growing body of work explores the use of AI for concept review, problem generation, and formative feedback (Trout & Winterbottom, 2025; Yeadon & Hardy, 2024), while other efforts describe the development of AI tutors, particularly those based on retrieval-augmented generation (RAG), to support active learning at scale (Kestin et al., 2024; Kortemeyer, 2024; Tufino, 2025). Although these contributions offer valuable insights, they often frame AI primarily as a neutral assistant rather than as a cognitive partner that both shapes and is shaped by students' thinking. Moreover, they rarely engage the ethical, epistemological, and metacognitive dimensions of AI use: How do students reason about appropriate use? What values inform their choices? How do these tools affect the way they learn and think?

The classroom activity described in this paper directly engages these questions. It provides instructors with a concrete, research-informed framework for inviting students into reflective, self-aware, and ethically grounded conversations about AI—conversations that are increasingly necessary as these tools become a routine part of academic life.

## Theoretical Foundations

Three foundational frameworks—social constructivist learning theory, metacognitive development, and ethics education—inform both the design and the intended outcomes of this activity.

Social constructivist theories of learning emphasize that understanding is actively constructed as learners interpret new ideas through the lens of prior experiences, knowledge, and ways of thinking, rather than passively absorbed (Bransford, Brown, & Cocking, 2000; Vygotsky, 1978). In physics education, this perspective is already familiar: research shows that meaningful learning typically emerges through activities that foster sustained student engagement, such as confronting preconceptions through research-based conceptual questions, grappling with challenging quantitative problems, and developing experimental and computational skills, particularly through dialogue and collaboration with peers and instructors within the classroom environment. Extending this logic to AI use, this activity engages students in analyzing familiar real-world scenarios, debating their implications, and co-constructing shared interpretations of acceptable practice, rather than simply being told what constitutes ethical or effective AI use. By inviting students to participate in the creation of classroom norms and policies, the activity positions them as partners in shaping the learning environment, an approach shown to increase student agency and sense of belonging (Cook-Sather, Bovill, & Felten, 2014).

Metacognition—the ability to reflect on and regulate one's own thinking—is likewise central to expert learning and scientific reasoning (Flavell, 1979). Students learn more effectively when they monitor how they are learning, assess the quality of their understanding, and evaluate whether their strategies are supporting meaningful progress. Generative AI introduces new challenges and opportunities for metacognitive growth by prompting students to reflect on both the answers they seek and how and why they are using these tools. This activity explicitly scaffolds that reflection by inviting students to evaluate both the ethical implications and learning value of AI use.

Finally, research in ethics education underscores the importance of reflective and dialogic approaches that encourage students to articulate, examine, and refine their ethical beliefs. Such approaches have been shown to deepen moral reasoning and support the long-term integration of ethical principles (Usher & Barak, 2024). In contrast to rule-based approaches to academic integrity, this activity invites students to wrestle with ambiguity and complexity, surface their own values, and engage in dialogue with peers. Ethical judgment is then framed as an ongoing practice within a learning community rather than a fixed set of rules.

Taken together, these frameworks provide a robust foundation for helping students navigate generative AI with curiosity, responsibility, and self-awareness. The goal is to cultivate the reflective habits that make ethical and effective learning possible.

## Engaging with the Practice

The activity presented below is intentionally designed to reflect the pedagogical frameworks outlined in the previous section. Drawing on social constructivist principles, it situates learning within shared dialogue and collaborative norm-setting rather than individual rule-following. In support of metacognitive development, the activity invites students to reflect on how and why they are using generative AI, not simply on the outputs it produces. Grounded in ethics education, the activity foregrounds ethical reasoning as a reflective and dialogic process rather than a set of fixed prescriptions.

Accordingly, rather than beginning with rules or prohibitions, the activity starts from authentic dilemmas—realistic scenarios of student interactions with AI—and uses these as a foundation for discussion, judgment, and the development of shared expectations about appropriate and effective AI use.

Students are presented with a set of 10–12 short scenarios involving the use (or non-use) of generative AI in a physics context. These examples span a range of practices, from ethically commendable uses that support conceptual understanding to clear violations of academic integrity. Working in small groups, students rank the scenarios from most ethical to least ethical and identify key boundaries—such as where AI use shifts from supporting learning to undermining it, and where it constitutes academic dishonesty. This exercise encourages nuanced ethical reasoning and metacognitive reflection, as students examine not only outcomes but also intent, understanding, and impact.

Following the ranking and discussion, students are invited to reflect on how their judgments align with institutional frameworks for academic integrity. Rather than treating these definitions as definitive answers, students consider where existing policies clarify, complicate, or fail to capture the distinctions they identified between AI use that supports learning and AI use that undermines it.

Building on this reflection, some implementations invite students to articulate shared guidelines or personal principles for AI use, often through individual reflection or writing. The goal is not simply to produce a policy document, but to cultivate ethical awareness, shared responsibility, and reflective engagement with learning.

This approach frames integrity as an active practice rather than a static rule. It positions students as thoughtful participants in a learning community evolving alongside new technologies and helps them develop critical distinctions between AI use that supports learning and AI use that circumvents it.

## Implementation

The activity consists of five core elements and several optional extensions. It typically takes about thirty minutes but can be extended through further discussion, written reflection, or policy work. It is adaptable to a variety of course formats and can be used early in the term, following an AI-supported assignment, or as part of a broader discussion on academic integrity and responsible technology use.[2]

A student-facing handout that guides students through each stage of the activity, including representative scenarios and reflection prompts, is provided as supplemental material.[3]

**Step 1: Understanding the Tools (5 minutes)**

The activity begins by broadening students' awareness of the AI landscape. Students are introduced to a range of AI tools and invited to reflect briefly on their prior familiarity before engaging in ethical evaluation.

**Step 2: Ranking Scenarios (10 minutes)**

Students are presented with 10–12 brief scenarios depicting varied uses of generative AI in a physics course. Working in small groups, students rank the scenarios from most to least ethically justifiable while articulating the reasoning behind their rankings. Prompts to guide discussion may include:

- How is the student using the AI?
- Does this use support or undermine the student's learning?
- What assumptions are you making about the student's goals or motivations, and how do those assumptions influence your judgment?

**Step 3: Identifying Key Boundaries (5 minutes)**

Each group is asked to reflect on their ranked list to identify one or more boundaries to indicate where important distinctions emerge. In particular, students are asked to identify where AI use shifts from supporting learning to undermining it, and where they believe AI use constitutes academic dishonesty. Groups may find that these boundaries coincide or fall in different places.

[2] In our case, we have successfully implemented the activity on the second day of class, which is typically when the first homework assignment is made available.
[3] For sample scenarios and printable classroom materials, please contact the author.

**Step 4: Connecting to Academic Integrity Frameworks (5 minutes)**

The instructor shares the institution's formal definition of academic integrity (or plagiarism). Students consider how this definition aligns with, complicates, or fails to capture the distinctions they identified in their discussions, including but not limited to differences in how AI use supports or undermines learning.

**Step 5: Class Discussion (10 minutes)**

The instructor facilitates a whole-class conversation comparing rankings and boundaries. Where is there consensus? Where do perspectives diverge? What ethical, epistemological, or practical principles are guiding students' evaluations? To what extent do students' goals for the course—such as learning the material versus earning a grade—influence how they draw boundaries around acceptable AI use? How do students distinguish between AI use that supports learning and AI use that constitutes academic dishonesty, and where do these boundaries align or diverge?

**Individual Reflection (Optional, Highly Recommended Extension)**

Students are invited to complete a brief written reflection after class, addressing prompts such as: What insights did you gain from the discussion? Did your thinking about AI use shift, and if so, how? How do you plan to use AI in ways that support your learning going forward? Submissions can be assessed on a pass/fail basis based on completion and thoughtfulness.

**Draft an AI Use Policy (Optional Individual or Collaborative Extension)**

As an additional extension, students draft a brief AI use policy for themselves, articulating how they plan to use—or intentionally avoid—AI tools in ways that support their learning and align with course goals. Students are encouraged to reflect on their own study habits, learning preferences, and areas of challenge. To support openness, this exercise is framed as non-evaluative, with no penalty for acknowledging past or current AI use, including practices students may now view as misaligned with their goals.

In smaller classes, this activity may be adapted for small-group policy drafting; in larger courses, individual policy writing may be more effective. Submissions can be assessed on a pass/fail basis based on completion and thoughtfulness.

**Revisiting the AI Use Policy (Optional End-of-Course Reflection)**

Students return to their original AI use policy near the end of the term and reflect on how their practices evolved. Prompts may include: Which aspects of your policy were easiest or hardest to follow? Where did your intentions align—or fail to align—with your actual study practices? How has your understanding of when AI supports or undermines learning changed over time? This return emphasizes ethical and constructive AI use as a developmental process rather than a one-time decision.

## Themes from Student Reflections

Following the activity, students report increased clarity not only about what constitutes appropriate use of AI, but also about how different forms of engagement with AI affect their learning. While individual responses vary, several themes consistently emerge across cohorts.

**AI as a Learning Tool and Metacognitive Awareness**

A central theme is a shift from viewing AI as a shortcut toward understanding it as a learning tool whose value depends on how it is used. Many students describe learning to distinguish between AI use that supports comprehension—such as prompting explanation, reflection, or self-testing—and use that bypasses intellectual effort. Students frequently connect this distinction to increased awareness of prompt design, noting that small changes in how a question is framed can significantly shape both the AI's response and their own engagement. Prompts such as "Ask me questions to help me understand this" are contrasted with more passive requests like "Explain how to solve this," leading students to reflect on how their phrasing influences the quality of their learning.

**Ethical Complexity and Value-Based Decision-Making**

The activity also surfaces ethical complexity in ways many students report encountering for the first time in a classroom setting. Rather than framing AI use as a simple binary of permissible or impermissible behavior, students begin to articulate the context-dependent nature of ethical decision-making. Many students note that this was the first course in which AI use had been explicitly acknowledged or discussed, and that this invitation to reflect—rather than ignore or prohibit—was both surprising and validating. Within this broader ethical reflection, some students raise concerns about the environmental impact of AI systems. A subset report intentionally limiting or avoiding AI use due to energy consumption and sustainability considerations, even when such tools might offer practical benefits. These reflections often reappear in students' individual AI use policies, where they describe navigating tensions between personal values and academic or social pressures.

**Ownership, Agency, and Ethical Identity**

Perhaps most importantly, students express a growing sense of ownership and confidence in their ethical decision-making. When invited to articulate and revisit their own AI use policies, students frame ethical judgment as something they actively practice rather than something imposed externally. This shift is captured succinctly in one student's comment during a late-quarter discussion: "I am really proud of how I used ChatGPT in this class." Rather than signaling compliance, such statements reflect a sense of agency, intentionality, and alignment between values and practice.

These themes are drawn from students' reflective writing and discussion rather than from formal assessment measures. While future implementations could support more systematic study of the activity's impact, the consistency and depth of students' responses suggest that it meaningfully engages students in reflective consideration of AI use.

Taken together, these reflections suggest that the activity fosters more than awareness of academic integrity policies. It supports metacognitive reflection, ethical reasoning, and epistemological maturity, while encouraging students to see themselves as active contributors to the ethical culture of their learning environments and disciplines.

## Conclusion: Co-Creating Integrity in the Age of AI

As generative AI becomes increasingly integrated into educational contexts, instructors face the dual challenge of maintaining academic integrity while supporting students in navigating new tools for learning. This activity offers one response: a structured, reflective framework that invites student reasoning, dialogue, and responsibility rather than prescriptive compliance.

Rather than positioning AI use as a threat to be constrained, this approach encourages students to examine how these tools can be used—or intentionally not used—in service of deep understanding. By analyzing scenarios, articulating boundaries around learning and integrity, and reflecting on their own practices over time, students engage in an active process of meaning-making that supports both metacognitive development and ethical growth.

The activity also models a broader pedagogical shift away from static policies and toward dynamic, participatory norm-setting. It reflects a view of academic integrity not as a fixed set of rules to be enforced, but as a culture to be cultivated—one rooted in shared values, reflective judgment, and intellectual honesty.

As instructors, we cannot fully predict how AI will continue to evolve. But we can equip students with habits of reflection and ethical reasoning that will allow them to engage with emerging technologies thoughtfully, responsibly, and in alignment with the values of the scientific and academic communities they are entering.